\documentstyle[prl,aps,psfig]{revtex} 
\begin{document}
\twocolumn[\hsize\textwidth\columnwidth\hsize\csname
@twocolumnfalse\endcsname 
\title{Low Mass Neutron Stars and the Equation of State of Dense Matter}
\author{J.~Carriere \footnote{e-mail: jcarrier@indiana.edu} and 
        C.~J.~Horowitz\footnote{e-mail:  charlie@iucf.indiana.edu}}
\address{Nuclear Theory Center and Dept. of Physics, Indiana
University, Bloomington, IN 47405}
\author{J.~Piekarewicz
\footnote{e-mail: jorgep@scri.fsu.edu}
}
\address{Department of Physics, Florida State University, 
Tallahassee, FL 32306}
\date{\today} 
\maketitle 
\begin{abstract}
Neutron-star radii provide useful information on the equation of state
of neutron rich matter. Particularly interesting is the density
dependence of the equation of state (EOS). For example, the softening
of the EOS at high density, where the pressure rises slower than
anticipated, could signal a transition to an exotic phase. However,
extracting the density dependence of the EOS requires measuring the
radii of neutron stars for a broad range of masses.  A ``normal'' $1.4
M_\odot$ ($M_\odot$=solar mass) neutron star has a central density of
a few times nuclear-matter saturation density ($\rho_0$). In contrast,
low mass ($\simeq\!0.5 M_\odot$) neutron stars have central densities
near $\rho_0$ so its radius provides information on the EOS at low
density.  Unfortunately, low-mass stars are rare because they may be
hard to form.  Instead, a precision measurement of nuclear radii on
atomic nuclei may contain similar information. Indeed, we find a
strong correlation between the neutron radius of $^{208}$Pb and the
radius of a $0.5 M_\odot$ neutron star. Thus, the radius of a $0.5
M_\odot$ neutron star can be inferred from a measurement of the the
neutron radius of $^{208}$Pb. Comparing this value to the measured
radius of a $\simeq\!1.4 M_\odot$ neutron star should provide the
strongest constraint to date on the density dependence of the equation
of state.
\end{abstract}
\vskip2.0pc]

\section{Introduction}
The structure of neutron stars, particularly their masses and radii, 
depend critically on the equation of state (EOS) of dense
matter~\cite{lattimer}. New measurements of masses and radii by 
state-of-the-art observatories should place important constraints on 
the EOS. Observing a rapid change of the EOS with density could signal 
a transition to an exotic phase of matter. Possibilities for new high 
density phases include pion or kaon condensates~\cite{kaon}, strange 
quark matter~\cite{quarks}, and/or a color 
superconductor~\cite{He01,Al01}. Measuring neutron-star radii $R(M)$ 
for a large range of neutron star masses $M$ is attractive as it
would allow one to directly deduce the EOS~\cite{Li92}, that is, the 
pressure as a function of the energy density $P(\epsilon)$.

While the masses of various neutron stars are accurately
known~\cite{masses}, precise measurements of their radii
do not yet exist. Therefore, several groups are devoting 
considerable effort at measuring neutron-star radii. Often 
one deduces the surface temperature $T_\infty$ and the 
luminosity $L$ of the star from spectral and distance 
measurements, respectively. Assuming a black-body spectrum, 
these measurements determine the surface area, and thus the 
effective radius $R_\infty$, of the star from the 
Stefan-Boltzmann law:
\begin{equation}
  L=4\pi\sigma R_\infty^2 T_\infty^4 \;.
\label{radius}
\end{equation}
Opportunities for precision measurements on neutron-star radii
include the isolated neutron star 
RX J185635-3754~\cite{walter,drake,pons} and
quiescent neutron stars in globular clusters, such as 
CXOU 132619.7-472910.8~\cite{bildsten}, where distances are
accurately known. Moreover, Sanwal and collaborators have recently 
detected absorption features in the radio-quiet neutron star 
1E~1207.4-5209~\cite{sanwal} that may provide the mass-to-radius 
ratio of the star through the determination of the gravitational 
redshift of the spectral lines. These observations are being 
complemented by studies that aim at constraining the composition 
of the neutron-star atmosphere~\cite{hailey}. Finally, models of 
rotational glitches place a lower limit on the radius of the Vela 
pulsar at $R_\infty\agt 12$~km~\cite{link}.

While a determination of the mass-radius relation $R(M)$ for a 
variety of neutron stars would uniquely determine the equation 
of state, unfortunately all accurately determined masses to date 
fall within a very small range. Indeed, a recent compilation by 
Thorsett and Chakrabarty of several radio binary pulsars place 
their masses in the narrow range of 
$1.25-1.44~M_{\odot}$~\cite{masses}. Note that several X-ray 
binaries appear to have larger masses, perhaps because of accretion.
These include Cyg~X-2 with a mass of $1.8\pm 0.2
M_\odot$~\cite{cygx2}, Vela X-1 with $1.9~M_\odot$~\cite{Ke95}, and 
4U~1700-37~\cite{He92}. If confirmed, they could provide additional 
information on the high density EOS. However, these mass
determinations are not without controversy~\cite{St97,Br96}. On the
other hand, it may be difficult to form low mass neutron stars from 
the collapse of heavier Chandrasekhar mass objects. If so, information 
on the low density EOS may not be directly available from neutron
stars. Thus, it is important to make maximum use of any $R(M)$
measurements even if these are available for only a limited range 
of masses.

Additional information on the low density EOS may be obtained from
precision measurements on atomic nuclei. For example, the neutron
radius of a heavy nucleus, such as $^{208}$Pb, is closely related to
the pressure of neutron rich matter~\cite{brown}. Indeed, heavy nuclei
develop a neutron-rich skin in response to this pressure.  The higher
the pressure the further the neutrons are pushed out against surface
tension, thereby generating a larger neutron radius.  However, nuclear
properties depend only on the EOS at normal (in the interior) and
below (in the surface) nuclear-matter saturation density
($\rho_0\approx 0.15$~nucleons/fm${}^{3}$). This is in contrast to
conventional 1.4~$M_\odot$ neutron stars that, with central densities
of several times $\rho_0$, also depend on the high-density component
of the EOS. This is not the case for low mass neutron stars (of about
1/2~$M_\odot$). Reaching central densities near $\rho_0$, {\it low
mass neutron stars probe the EOS at similar densities as atomic
nuclei.} Therefore, one could infer the radius of low mass neutron
stars from detailed measurements on atomic nuclei.  Having inferred
the radii of low mass neutron stars from a nuclear measurement,
combined with the measured radius of a $1.4 M_\odot$ star, may enable
one to deduce the density dependence of the EOS.  (For a recent
discussion on the minimum stable mass of a neutron star see
Ref.\cite{min}.)  The parity radius experiment at Jefferson Laboratory
\cite{prex} aims to measure accurately and model independently the
root-mean-square neutron radius ($R_n$) of $^{208}$Pb via parity
violating elastic electron scattering~\cite{bigpaper}. Such an
experiment probes neutron densities because the weak vector charge of
a neutron is much larger than that of a proton. The goal of the
experiment is to measure $R_n$ to a 1\% accuracy (within
$\approx\pm0.05$~fm). 

The outline of the paper is as follows. In Sec.~\ref{sec:formalism} we
present a relativistic effective field theory formalism to study
relationships between the neutron radius of $^{208}$Pb and the radii
of neutron stars. Uncertainties in these relationships are estimated
by considering a wide range of effective field theory parameters, all
of them constrained by known nuclear properties. This formalism has
been used previously to study correlations between the neutron radius 
of $^{208}$Pb and the properties of the neutron-star crust~\cite{prl}, 
the radii of 1.4~$M_\odot$ neutron stars~\cite{radii}, and the direct 
URCA cooling of neutron stars~\cite{hpurca}.
As the radius of low mass neutron stars depends on the solid crust
of nonuniform matter, a treatment of the crust is discussed in
Sec.~\ref{sec:crust}. Our results, presented in
Sec.~\ref{sec:results}, show a strong correlation between the radius 
of a low mass neutron star and the neutron radius $R_n$ of $^{208}$Pb
that is essentially model independent. This is because the structure 
of both objects depend on the EOS at similar densities. In contrast, 
the radius of a 1.4 $M_\odot$ neutron star shows a considerable model 
dependence. This is because a 1.4~$M_\odot$ neutron star is also
sensitive to the EOS at higher densities, and the high density EOS is 
not constrained by nuclear observables. In Sec.~\ref{sec:conclusions}
we conclude that properties of low mass neutron stars can be inferred 
from measuring properties of atomic nuclei. In particular, the radius 
of a 1/2 $M_\odot$ neutron star can be deduced from a measurement of 
$R_n$ in $^{208}$Pb. One will then be able to directly compare this
inferred radius to the measured radius of an $\approx 1.4~M_\odot$
neutron star to gain information on the density dependence of the EOS.
Thus, even if a mass-radius measurement for a single ($\approx
1.4~M_\odot$) neutron star is available, one can use the atomic
nucleus to gain information on the density dependence of the EOS. This
should provide the most precise determination of the density
dependence of the EOS to date and should indicate whether a transition
to a high density exotic phase of matter is possible or not.

\section{Formalism}
\label{sec:formalism}
Our starting point is the relativistic effective-field theory of
Ref.~\cite{horst} supplemented with new couplings between the
isoscalar and the isovector mesons. This allows us to correlate
nuclear observables, such as the neutron radius of $^{208}$Pb,  
with neutron star properties. We will explore uncertainties in 
these correlations by considering a range of model parameters. 
The model has been introduced and discussed in detail in several 
earlier references~\cite{prl,radii,hpurca}, yet a brief summary is 
included here for completeness.

The interacting Lagrangian density is given by~\cite{prl,horst}
\begin{eqnarray}
{\cal L}_{\rm int} &=&
\bar\psi \left[g_{\rm s}\phi   \!-\! 
         \left(g_{\rm v}V_\mu  \!+\!
    \frac{g_{\rho}}{2}{\mbox{\boldmath $\tau$}}\cdot{\bf b}_{\mu} 
                               \!+\!    
    \frac{e}{2}(1\!+\!\tau_{3})A_{\mu}\right)\gamma^{\mu}
         \right]\psi \nonumber \\
                   &-& 
    \frac{\kappa}{3!} (g_{\rm s}\phi)^3 \!-\!
    \frac{\lambda}{4!}(g_{\rm s}\phi)^4 \!+\!
    \frac{\zeta}{4!}   g_{\rm v}^4(V_{\mu}V^\mu)^2 
    \nonumber \\
                   &+&
    g_{\rho}^{2}\,{\bf b}_{\mu}\cdot{\bf b}^{\mu}
    \left[\Lambda_{\rm s} g_{\rm s}^{2}\phi^2 +
          \Lambda_{\rm v} g_{\rm v}^{2}V_{\mu}V^\mu\right] \;.
 \label{LDensity}
\end{eqnarray}
The model contains an isodoublet nucleon field ($\psi$) interacting
via the exchange of two isoscalar mesons, the scalar sigma ($\phi$)and
the vector omega ($V^{\mu}$), one isovector meson, the rho 
(${\bf b}^{\mu}$), and the photon ($A^{\mu}$). In addition to meson-nucleon
interactions the Lagrangian density includes scalar and vector
self-interactions.  Omega-meson self-interactions $\zeta$ soften the
equation of state at high density.  Finally, the nonlinear couplings
$\Lambda_{\rm s}$ and $\Lambda_{\rm v}$ are included to modify the
density-dependence of the symmetry energy 
$a_{\rm sym}(\rho)$~\cite{prl,radii,hpurca}. We employ
Eq.~(\ref{LDensity}) in a mean field approximation where the meson 
fields are replaced by their ground state expectation values.
The couplings constants in Eq. (\ref{LDensity}) are fit to nuclear 
matter and finite nuclei properties.  All of the parameter sets 
considered here, namely, NL3~\cite{NL3}, S271~\cite{prl}, and 
Z271~\cite{prl} reproduce the following properties of symmetric 
nuclear matter: saturation at a Fermi momentum of 
$k_F=1.30~$fm$^{-1}$ with a binding energy per nucleon of 
$-16.24$~MeV and an incompressibility of $K=271$~MeV. The various
parameter sets differ in their effective masses at saturation density,
in their $\omega$-meson self interactions (which are included for Z271 
and neglected for NL3 and S271) and in the nonlinear couplings 
$\Lambda_{\rm s}$ and $\Lambda_{\rm v}$ (see Table~\ref{Tableone}).
Note that the NL3 parametrization has been used extensively to
reproduce a variety of nuclear properties~\cite{NL3}.

The symmetry energy at saturation density is not well constrained
experimentally. However, an average of the symmetry energy at
saturation density and the surface symmetry energy is constrained by
the binding energy of nuclei. Thus, the following prescription has
been adopted: the value of the $NN\rho$ coupling constant is adjusted
so that all parameter sets have a symmetry energy of $25.67$~MeV at
$k_F\!=\!1.15$~fm$^{-1}$. This insures accurate binding energies for
heavy nuclei, such as $^{208}$Pb. Following this prescription the
symmetry energy at saturation density is predicted to be $37.3$,
$36.6$, and $36.3$~MeV for parameter sets NL3, S271, and Z271,
respectively (for $\Lambda_{\rm s}\!=\!\Lambda_{\rm v}\!=\!0$).
Changing $\Lambda_{\rm s}$ or $\Lambda_{\rm v}$ changes the density
dependence of the symmetry energy by changing the effective rho-meson
mass. In general increasing either $\Lambda_{\rm s}$ or $\Lambda_{\rm
v}$ causes the symmetry energy to grow more slowly with density.

The neutron radius of $^{208}$Pb depends on the density dependence of
the symmetry energy.  A large pressure for neutron matter pushes
neutrons out against surface tension and leads to a large neutron
radius.  The pressure depends on the derivative of the energy of
symmetric matter with respect to density (which is approximately
known) and the derivative of the symmetry energy, $da_{\rm
sym}/d\rho$. Thus parameter sets with a large $da_{\rm sym}/d\rho$
yield a large neutron radius in $^{208}$Pb.  Note that all parameter
sets approximately reproduce the observed proton radius and binding
energy of $^{208}$Pb.  Therefore changing $\Lambda_{\rm s}$ or
$\Lambda_{\rm v}$ allows one to change the density dependence of the
symmetry energy $da_{\rm sym}/d\rho$, while keeping many other
properties fixed. Once the model parameters have been fixed, it is a 
simple matter to calculate the EOS for uniform matter in beta
equilibrium, where the chemical potentials of the neutrons $\mu_n$, 
protons $\mu_p$, electrons $\mu_e$, and muons $\mu_\mu$ satisfy,
\begin{equation}
\mu_n-\mu_p=\mu_e=\mu_\mu\;.
\label{betaeq}
\end{equation}
Note that the high density interior of a neutron star is assumed to
be a uniform liquid; possible transitions to a quark- or
meson-condensate phase are neglected.   

\begin{table}
\caption{Model parameters used in the calculations. The 
parameter $\kappa$ and the scalar mass $m_{\rm s}$ are 
given in MeV. The nucleon, rho, and omega masses are kept 
fixed at $M\!=\!939$, $m_{\rho}\!=\!763$, and 
$m_{\omega}\!=\!783$~MeV, respectively --- except in the 
case of the NL3 model where it is fixed at 
$m_{\omega}\!=\!782.5$~MeV.}
\begin{tabular}{lcccccc}
 Model & $m_{\rm s}$  & $g_{\rm s}^2$ & $g_{\rm v}^2$ & 
           $\kappa$ & $\lambda$ & $\zeta$ \\
 NL3  & 508.194 & 104.3871 & 165.5854 & 3.8599 & $-$0.0159049 & 0 \\ 
 S271 & 505     &  81.1071 & 116.7655  & 6.68344 & $-$0.01580 & 0 \\ 
 Z271 & 465     &  49.4401 &  70.6689  & 6.16960 & $+$0.156341 & \ 0.06     
\label{Tableone}
\end{tabular}
\end{table}

\section{Boundary Between Crust and Interior}
\label{sec:crust}
Neutron stars are expected to have a solid inner crust of nonuniform
neutron-rich matter above a liquid mantle. The phase transition from
solid to liquid is thought to be weakly first order and can be found
by comparing a detailed model of the nonuniform crust to the liquid
(see for example~\cite{dh1}). Yet in practice, model calculations
yield very small density discontinuities at the transition. Therefore
a good approximation is to search for the density where the uniform
liquid first becomes unstable to small amplitude density oscillations
(see for example \cite{dh2}). This method would yield the exact
transition density for a second order phase transition.

The stability analysis of the uniform ground state is based on
the relativistic random-phase-approximation (RPA) of 
Ref.~\cite{cjhkw} for a system of electrons, protons, and neutrons. 
The approach is generalized here to accommodate the various nonlinear
couplings among the meson fields. We start by considering a plane wave
density fluctuation of momentum $q=|{\bf q}|$ and zero energy
$q_0\!=\!0$. To describe small amplitude particle-hole (or
particle-antiparticle) excitations of the fermions we compute the
longitudinal polarization matrix that is defined as follows:
\begin{equation}
\Pi_L=\pmatrix{\Pi_{00}^e&0&0&0\cr 0&\Pi_s^n+\Pi_s^p&\Pi_m^p&\Pi_m^n\cr
0&\Pi_m^p&\Pi_{00}^p&0\cr 0&\Pi_m^n&0&\Pi_{00}^n}\;.
\label{Pi}
\end{equation}
Here the one-one entry describes electrons, the two-two entry protons
plus neutrons interacting via scalar mesons, the three-three entry
protons interacting with vector mesons and the four-four entry
neutrons interacting with vector mesons. The individual polarization 
insertions are given by
\begin{mathletters}
\begin{eqnarray}
i\Pi_s(q,q_0)&=&\int {d^4p\over (2\pi)^4} 
{\rm Tr}\Big[G(p)G(p+q)\Big] \;, \label{pis} \\
i\Pi_m(q,q_0)&=&\int {d^4p\over (2\pi)^4} 
{\rm Tr}\Big[G(p)\gamma_0 G(p+q)\Big] \;, \label{pim} \\
i\Pi_{00}(q,q_0)&=&\int {d^4p\over (2\pi)^4} 
{\rm Tr}\Big[G(p)\gamma_0 G(p+q) \gamma_0\Big]\;,
\label{pil}
\end{eqnarray}
\end{mathletters}
where ${\rm Tr}$ indicates a trace over Dirac indices. Note that the
fermion Green's function has been defined as
\begin{equation}
G(p)=(\slash\hskip -5.1pt p + M^*)
 \left({1\over {p^*}^2-{M^*}^2} + 
 {i\pi\over E_p^*}\delta(p_0^*-E_p^*)\theta(k_F-|{\bf p}|)\right) \;.
\end{equation}
Here $k_F$ is the Fermi momentum, $M^*=M\!-\!g_s\phi_0$ is the nucleon 
effective mass, $E_p^*=(p^2+{M^*}^2)^{1/2}$, and 
$p_\mu^*=p_\mu-(g_vV_\mu\pm g_\rho b_\mu/2)$ (with the plus sign 
for protons and the minus sign for neutrons). Note that in the case
of the electrons $M^*=m_e$ and $p_\mu^*=p_\mu$. Explicit analytic 
formulas for $\Pi_{00}$, $\Pi_s$, and $\Pi_m$ in the static limit 
($q_0\!=\!0$) are given in the appendix.

The lowest order meson propagator $D_L^0$ is computed in
Ref.~\cite{cjhkw} in the absence of nonlinear meson couplings.
It is given by,
\begin{equation}
    D_L^0 = \pmatrix{d_g&0&-d_g&0\cr 
                     0&-d_{\rm s}^0 & 0 & 0\cr 
                    -d_g &0 & d_g+d_{\rm v}^0+d_\rho^0& 
		     d_{\rm v}^0-d_\rho^0\cr
                     0&0&d_{\rm v}^0-d_\rho^0&d_{\rm v}^0+d_\rho^0}\;.
\label{DLong0}
\end{equation}
Expressions for the photon and for the various meson propagators
in the limit of no nonlinear meson couplings are given as follows:
\begin{mathletters} 
\begin{eqnarray}
  d_{g}&=&\frac{e^2}{q^2}=\frac{4\pi\alpha}{q^2} \;, \\
  d_{\rm s}^{0}&=&\frac{g_{\rm s}^{2}}{q^2+m_{\rm s}^2} \;, \\
  d_{\rm v}^{0}&=&\frac{g_{\rm v}^{2}}{q^2+m_{\rm v}^2} \;, \\
  d_\rho^0&=&\frac{g_{\rho}^{2}/4}{q^2+m_{\rho}^2} \;.
\end{eqnarray}
\label{DLong}
\end{mathletters}
The appearance of a minus sign in the one-three element of $D_L^0$ 
relative to the one-one element is because electrons and protons 
have opposite electric charges.

The addition of nonlinear couplings in the Lagrangian leads to a 
modification of the meson masses. Effective meson masses are 
defined in terms of the quadratic fluctuations of the meson fields 
around their static, mean-field values (the linear fluctuations
vanish by virtue of the mean-field equations). That is,
\begin{equation}
  m_{\rm s}^{*2}=-\frac{\partial^{2}{\cal L}}{\partial\phi_0^{2}}\;,
   \quad
  m_{\rm v}^{*2}=+\frac{\partial^{2}{\cal L}}{\partial V_0^{2}}\;,
   \quad
  m_{\rho }^{*2}=+\frac{\partial^{2}{\cal L}}{\partial b_0^{2}}\;.
\end{equation}
This yields the following expressions for the effective meson masses 
in terms of the static meson fields and the coupling constants defined
in the interacting Lagrangian of Eq.~(\ref{LDensity}):
\begin{mathletters}
\begin{eqnarray}
  m_{\rm s}^{*2}&=&m_{\rm s}^2+g_{\rm s}^2
     \left(\kappa \Phi_0 + {\lambda\over 2} 
     \Phi_0^2-2\Lambda_{\rm s} B_0^2\right)\;, \\
  m_{\rm v}^{*2}&=&m_{\rm v}^2+g_{\rm v}^2
     \left({\zeta\over 2}W_{0}^{2}+ 
     2\Lambda_{\rm v} B_0^2\right)\;, \\
  m_{\rho}^{*2}&=&m_{\rho}^2+g_{\rho}^2
     \left(2\Lambda_{\rm s} \Phi_0^2+
           2\Lambda_{\rm v} W_0^2\right)\;. 
\end{eqnarray}
\end{mathletters}
Note that the following definitions have been introduced:
$\Phi_{0}\!\equiv\!g_{\rm s}\phi_{0}$, 
$W_{0}\!\equiv\!g_{\rm v}V_{0}$, and
$B_{0}\!\equiv\!g_{\rho}b_{0}$.

Further, the new couplings between isoscalar and isovector mesons 
($\Lambda_{\rm s}$ and $\Lambda_{\rm v}$) lead to additional off 
diagonal terms in the meson propagator. These arise because the
quadratic fluctuations around the static solutions generate terms
of the form 
\begin{equation}
  \frac{\partial^{2}{\cal L}}{\partial\phi_0\partial b_0}\neq0
  \quad {\rm and} \quad
  \frac{\partial^{2}{\cal L}}{\partial V_0\partial b_0}\neq0 \;.
\end{equation}
For simplicity we only consider here the following two cases: 
{\it i)}  ($\Lambda_{\rm s}\neq 0$ and $\Lambda_{\rm v}\!=\!0$) or 
{\it ii)} ($\Lambda_{\rm s}\!=\!0$ and $\Lambda_{\rm v}\neq 0$),
and neglect the (slightly) more complicated case in which both
coupling constants are different from zero.

For the first case of ($\Lambda_{\rm s}\neq 0$ and 
$\Lambda_{\rm v}\!=\!0$) the new components of the
longitudinal meson propagator become
\begin{mathletters}
\begin{eqnarray}
  d_{\rm v}&=&\frac{g_{\rm v}^{2}}
               {q^2+m_{\rm v}^{*2}}\;, \\
  d_{\rm s}&=&\frac{g_{\rm s}^{2}(q^2+m_\rho^{*2})}
               {(q^2+m_{\rm s}^{*2})(q^2+m_\rho^{*2})+
               (4g_{\rm s}g_{\rho}\Lambda_{\rm s}
	       \Phi_{0}B_{0})^{2}} \;, \\
  d_{\rho}&=&\frac{(g_{\rho}^{2}/4)(q^2+m_{\rm s}^{*2})}
               {(q^2+m_{\rm s}^{*2})(q^2+m_\rho^{*2})+
               (4g_{\rm s}g_{\rho}\Lambda_{\rm s}
	       \Phi_{0}B_{0})^{2}} \;, \\
  d_{{\rm s}\rho}&=&
               \frac{2g_{\rm s}^{2}g_{\rho}^{2}
	       \Lambda_{\rm s}\Phi_{0}B_{0}}
               {(q^2+m_{\rm s}^{*2})(q^2+m_\rho^{*2})+
               (4g_{\rm s}g_{\rho}\Lambda_{\rm s}
	       \Phi_{0}B_{0})^{2}} \;.
\end{eqnarray} 
\label{DLongsr}
\end{mathletters}
With these changes the modified longitudinal meson propagator now reads,
\begin{equation}
D_L = \pmatrix{d_g&0&-d_g&0\cr 
               0&-d_{\rm s} & d_{{\rm s}\rho} & -d_{{\rm s}\rho}\cr 
              -d_g &d_{{\rm s}\rho} & d_g+d_{\rm v}+
	       d_\rho& d_{\rm v}-d_\rho\cr
               0&-d_{{\rm s}\rho}&d_{\rm v}-d_\rho&d_{\rm v}+d_\rho}.
\end{equation}

Alternatively, for the case of 
($\Lambda_{\rm s}\!=\!0$ and $\Lambda_{\rm v}\neq 0$) we obtain,
\begin{mathletters}
\begin{eqnarray}
  d_{\rm s}&=&\frac{g_{\rm s}^{2}}
               {q^2+m_{\rm s}^{*2}}\;, \\
  d_{\rm v}&=&\frac{g_{\rm v}^{2}(q^2+m_\rho^{*2})}
               {(q^2+m_{\rm v}^{*2})(q^2+m_\rho^{*2})-
               (4g_{\rm v}g_{\rho}\Lambda_{\rm v}
	       W_{0}B_{0})^{2}} \;, \\
  d_{\rho}&=&\frac{(g_{\rho}^{2}/4)(q^2+m_{\rm v}^{*2})}
               {(q^2+m_{\rm v}^{*2})(q^2+m_\rho^{*2})-
               (4g_{\rm v}g_{\rho}\Lambda_{\rm v}
	       W_{0}B_{0})^{2}} \;, \\
  d_{{\rm v}\rho}&=&
               \frac{-2g_{\rm v}^{2}g_{\rho}^{2}
	       \Lambda_{\rm v}W_{0}B_{0}}
               {(q^2+m_{\rm v}^{*2})(q^2+m_\rho^{*2})-
               (4g_{\rm v}g_{\rho}\Lambda_{\rm v}
	       W_{0}B_{0})^{2}} \;.
\end{eqnarray} 
\label{DLongvr}
\end{mathletters}

With these changes the modified longitudinal meson propagator now reads,
\begin{equation}
D_L = \pmatrix{d_g&0&-d_g&0\cr 
               0&-d_{\rm s} & 0 & 0\cr -d_g & 0 & 
	       d_g\!+\!d_{\rm v}\!+\!d_\rho\!+\!2d_{{\rm v}\rho}& 
               d_{\rm v}\!-\!d_\rho\cr
	       0&0&d_{\rm v}\!-\!d_\rho&d_{\rm v}\!+\!d_\rho\!-\!
	       2d_{{\rm v}\rho}}.
\end{equation}

\begin{figure}
\psfig{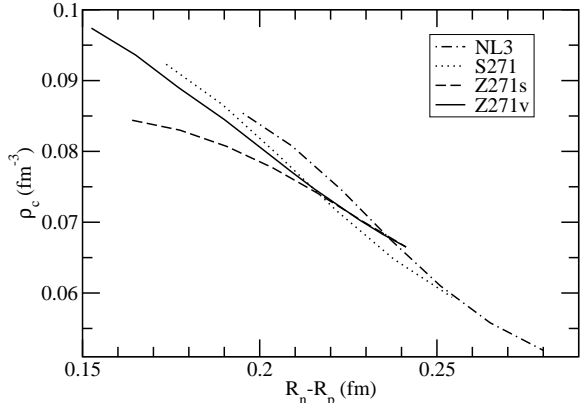}
\caption{The transition density $\rho_c$ at which uniform matter 
         becomes unstable to density oscillations as a function of 
         the neutron skin in $^{208}$Pb. The solid line is for the 
	 Z271 parameter set with $\Lambda_{\rm v}\neq 0$ while the 
 	 dashed curve uses Z271 with $\Lambda_{\rm s}\neq 0$. The 
	 dotted curve is for the S271 set and the dot-dashed curve 
	 for NL3, both of these with $\Lambda_{\rm v}\neq 0$.}
\label{Fig1}
\end{figure}

The uniform system becomes unstable to small amplitude density 
fluctuations of momentum transfer $q$ when the following condition 
is satisfied:
\begin{equation}
{\rm det} \Big[1 - D_L(q) \Pi_L(q,q_0\!=\!0) \Big] \leq 0\;.
\label{det}
\end{equation}
We estimate the transition density ($\rho_c$) between the inner 
crust and the liquid interior as the largest density for which
Eq.~(\ref{det}) has a solution. Our results for $\rho_c$ are listed 
in Tables~\ref{Tabletwo}-\ref{Tablefive} and also shown in 
Fig.~\ref{Fig1}. We find a strong correlation between the 
neutron skin of $^{208}$Pb and $\rho_c$, as originally 
discussed in Ref.~\cite{prl}.
At the lower densities of the inner crust the system is nonuniform 
and may have a very complex structure that may include spherical, 
cylindrical, and plate-like nuclei, bubbles, rods, plates, 
{\it etc.}~\cite{Lo93,Oy93}. At present we do not have microscopic 
calculations of these structures in our models. Therefore we adopt 
a simple interpolation formula to estimate the equation of state in
the inner crust. That is, we assume a polytropic form for the EOS 
in which the pressure is approximately given by~\cite{link},
\begin{equation}
 P(\epsilon)=A+B\epsilon^{4/3},
\label{ab}
\end{equation}
where $\epsilon$ is the mass-energy density. The two constants $A$ 
and $B$ in Eq.~(\ref{ab}) are chosen so that the pressure is
continuous at the boundary between the inner crust and the
liquid interior (determined from the RPA analysis) and at the 
boundary between the inner and the outer crusts. For the low 
density outer crust we assume the EOS of Baym, Pethick, and 
Sutherland(BPS)~\cite{baym} up to a baryon density of 
$\rho_{\rm outer}\!=\!2.57\times 10^{-4}$~fm$^{-3}$ which
corresponds to an energy density of 
$\epsilon_{\rm outer}\!=\!~4.30\times 10^{11}$~g/cm$^3$ 
(or 0.24~MeV/fm$^3$) and a pressure of 
$P_{\rm outer}\!=\!4.87\times 10^{-4}$~MeV/fm$^3$.
Thus the two constants $A$ and $B$ of Eq.~(\ref{ab}) are adjusted
to reproduce $P_{\rm outer}$ at $\epsilon_{\rm outer}$ and the pressure 
of the uniform liquid, calculated within the relativistic mean-field
(RMF) approach, at $\epsilon_c$ which is the energy density
corresponding to $\rho_c$. That is,
\begin{equation}
  P(\epsilon)=\cases{
    P_{\rm BPS}(\epsilon)\;, & 
    for $\epsilon_{\rm min}\le\epsilon\le\epsilon_{\rm outer}\;;$ \cr
    A+B\epsilon^{4/3} \;, &
    for $\epsilon_{\rm outer}<\epsilon\le\epsilon_{c}\;;$ \cr
    P_{\rm RMF}(\epsilon)\;, & 
    for $\epsilon_{c}<\epsilon \;.$}
\end{equation}
Note that $\epsilon_{\rm min}\!=\!5.86\times 10^{-9}$~MeV/fm$^3$
is the minimum value of the energy density included in the equation 
of state. This value corresponds to a minimum pressure of 
$P_{\rm min}\!=\!6.08\times 10^{-15}$~MeV/fm$^3$, which is the 
value at which we stop integrating the Tolman-Oppenheimer-Volkoff 
equations. That is, the radius $R$ of a neutron star 
(see Sec.~\ref{sec:results}) is defined by the expression
$P(R)\!=\!P_{\rm min}$. For the relativistic mean field interaction 
(TM1) of Ref.~\cite{tm1}, the relatively simple procedure presented
here is a good approximation to the more complicated explicit 
calculation of the EOS in the inner crust~\cite{tm1eos}.

\section{Results}
\label{sec:results}

\begin{figure}[h]
\psfig{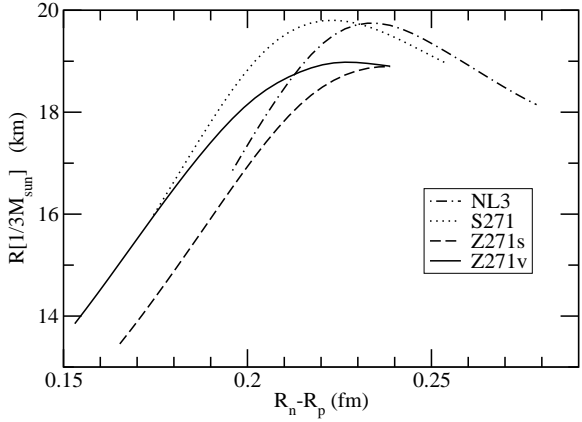}
\caption{Radius of a neutron star of mass 1/3 $M_\odot$ 
         as a function of the neutron skin in $^{208}$Pb. 
	 The solid line is for the Z271 parameter set with 
	 $\Lambda_{\rm v}\neq 0$ while the dashed curve 
	 uses Z271 with $\Lambda_{\rm s}\neq 0$. The 
	 dotted curve is for the S271 set and the 
	 dot-dashed curve for NL3, both of these with 
	 $\Lambda_{\rm v}\neq 0$.}
\label{Fig2}
\end{figure}

\begin{figure}[h]
\psfig{file=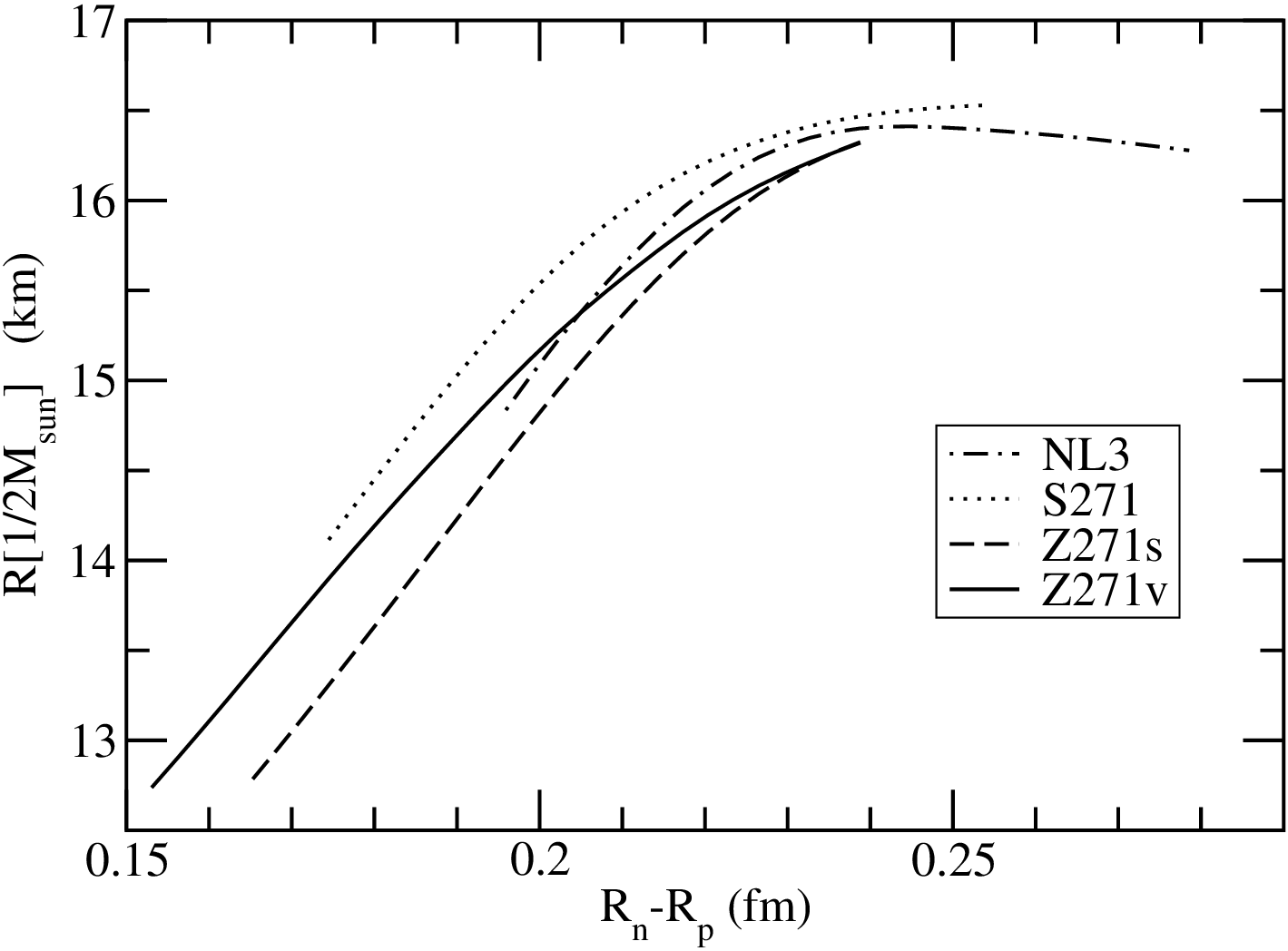,width=3in}
\caption{Radius of a neutron star of mass 1/2 $M_\odot$ 
         as a function of the neutron skin in $^{208}$Pb. 
	 The solid line is for the Z271 parameter set with 
	 $\Lambda_{\rm v}\neq 0$ while the dashed curve 
	 uses Z271 with $\Lambda_{\rm s}\neq 0$. The 
	 dotted curve is for the S271 set and the 
	 dot-dashed curve for NL3, both of these with 
	 $\Lambda_{\rm v}\neq 0$.}
\label{Fig3}
\end{figure}

\begin{figure}[h]
\psfig{file=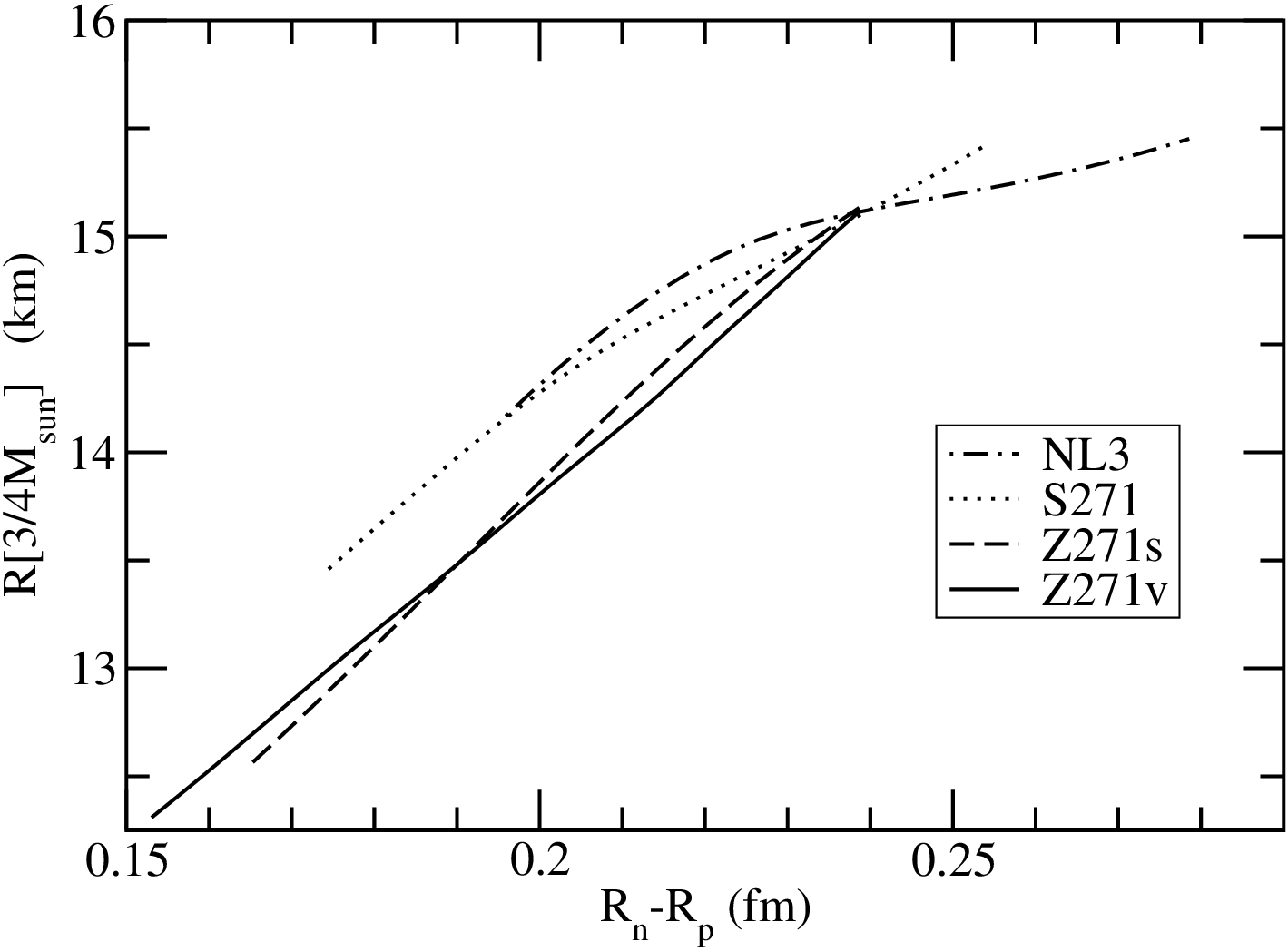,width=3in}
\caption{Radius of a neutron star of mass 3/4 $M_\odot$ 
         as a function of the neutron skin in $^{208}$Pb. 
	 The solid line is for the Z271 parameter set with 
	 $\Lambda_{\rm v}\neq 0$ while the dashed curve 
	 uses Z271 with $\Lambda_{\rm s}\neq 0$. The 
	 dotted curve is for the S271 set and the 
	 dot-dashed curve for NL3, both of these with 
	 $\Lambda_{\rm v}\neq 0$.}
\label{Fig4}
\end{figure}

\begin{figure}[h]
\psfig{file=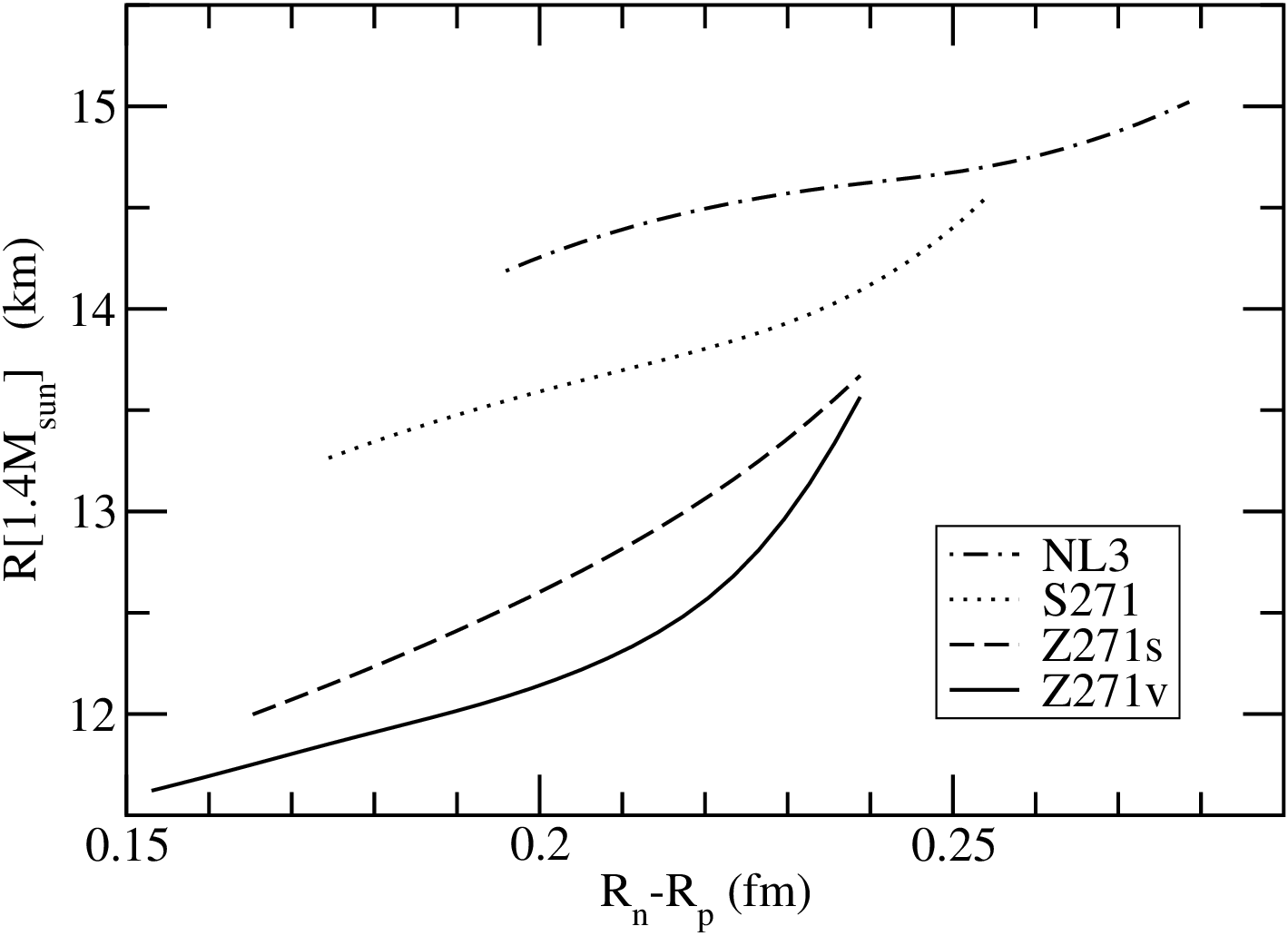,width=3in}
\caption{Radius of a neutron star of mass 1.4 $M_\odot$ 
         as a function of the neutron skin in $^{208}$Pb. 
	 The solid line is for the Z271 parameter set with 
	 $\Lambda_{\rm v}\neq 0$ while the dashed curve 
	 uses Z271 with $\Lambda_{\rm s}\neq 0$. The 
	 dotted curve is for the S271 set and the 
	 dot-dashed curve for NL3, both of these with 
	 $\Lambda_{\rm v}\neq 0$.}
\label{Fig5}
\end{figure}

Figures~\ref{Fig2}-\ref{Fig5} and
Tables~\ref{Tabletwo}-\ref{Tablefive} 
show the radii of neutron stars of mass $1/3, 1/2, 3/4$ and
$1.4~M_\odot$ as a function of the neutron skin ($R_n$-$R_p$) of
$^{208}$Pb.  One might expect a strong correlation between the radius
of a neutron star and the neutron radius of $^{208}$Pb, as the same
pressure of neutron rich matter that pushes neutrons out against
surface tension in $^{208}$Pb pushes neutrons out against gravity in a
neutron star~\cite{radii}. However, the central density of a
$1.4M_\odot$ neutron star is a few times larger than normal
nuclear-matter saturation density $\rho_0$. Thus, $R(1.4M_\odot)$
depends on the EOS at low and high densities while $R_n$ only depends
on the EOS for $\rho\!\leq\!\rho_0$.  Softening the EOS ({\it i.e.,}
decreasing the pressure) at high densities will decrease
$R(1.4M_\odot)$ without changing $R_n$. Hence, while Fig.~\ref{Fig5}
shows a definite correlation---$R(1.4M_\odot)$ grows with increasing
$R_n$-$R_p$---there is a strong model dependence.

In contrast, the central density of a $\frac{1}{2}M_\odot$ star is of
the order of $\rho_0$, so $R(\frac{1}{2}M_\odot)$ and $R_n$ depend on
the EOS over a comparable density range. As a result, we find a strong
correlation and weak model dependence in Fig.~\ref{Fig3}. For example,
if $R_n$-$R_p$ in $^{208}$Pb is relatively large, {\it e.g,}
$R_n\!-\!R_p\!\approx\!0.25$~fm, then
$R(\frac{1}{2}M_\odot)\!\approx\!16$~km.  Alternatively, if
$R_n\!-\!R_p\!\approx\!0.15$~fm, then 
$R(\frac{1}{2}M_\odot)\alt 13$~km. This is an important result. It 
suggest that even if observations of low mass neutron stars are not 
feasible, one could still infer their radii from a single nuclear 
measurement. Note that the results for a $\frac{3}{4}M_\odot$ neutron 
star (Fig.~\ref{Fig4}) follow a similar trend. 

We conclude this section with a comment on $\frac{1}{3}M_\odot$
neutron stars. Parameter sets that generate very large values
for $R_n\!-\!R_p$ have large pressures near $\rho_0$. This implies 
that the energy of neutron rich matter rises rapidly with density. 
In turn, this leads (because all parameter sets are constrained to 
have the same symmetry energy at $\rho\!=\!0.1$~fm$^{-3}$) to lower 
energies and {\it lower pressures} at very low density as compared 
with parameter sets with smaller values for $R_n\!-\!R_p$. This 
low-density region is important for low mass neutron stars and could 
explain why $R({1\over 3}M_\odot)$ in Fig.~\ref{Fig2} actually 
decreases with increasing neutron skin for $R_n\!-\!R_p\agt 0.23$~fm.

\twocolumn[\hsize\textwidth\columnwidth\hsize\csname
@twocolumnfalse\endcsname 

\begin{table}
\caption{Results for the NL3 parameter set with 
         $\Lambda_{\rm s}\!=\!0$. The $NN\rho$ coupling 
         constant $g_\rho^2$ and the neutron minus proton 
	 root mean square radius for $^{208}$Pb (in fm) 
	 are given along with the radii of 1/3, 1/2, 3/4 
	 and 1.4 $M_\odot$ neutron stars in km. Finally, 
	 the transition density $\rho_c$ between the inner 
	 crust and liquid interior is given in fm$^{-3}$.}
\begin{tabular} {lccccccc}
$\Lambda_{\rm v}$ & $g_\rho^2$ & $R_n-R_p$($^{208}$Pb) 
& R(${1\over 3}M_\odot$) & R(${1\over 2}M_\odot$) 
& R(${3\over 4}M_\odot$) & R($1.4M_\odot$) & $\rho_c$ \\
0.030 & 127.0 & 0.1952 & 16.766 & 14.789 & 14.142 & 14.175& 0.0854 \\
0.025 & 115.6 & 0.209  & 18.37 & 15.59 & 14.60 & 14.38 & 0.0808 \\
0.020 & 106.0 & 0.223  & 19.49 & 16.15 & 14.93 & 14.52 & 0.0746 \\
0.015 & 97.9 & 0.237  & 19.73 & 16.39 & 15.10 & 14.61 & 0.0675 \\
0.010 & 90.9 & 0.251  & 19.31 & 16.40 & 15.20 & 14.68 & 0.0610 \\
0.005 & 84.9 & 0.265  & 18.70 & 16.35 & 15.31 & 14.81 & 0.0558 \\
0.000 & 79.6 & 0.280  & 18.10 & 16.27 & 15.47 & 15.05 & 0.0519 \\
\label{Tabletwo}
\end{tabular}
\end{table}

\begin{table}
\caption{Results for the S271 parameter set with 
         $\Lambda_{\rm s}\!=\!0$. The $NN\rho$ coupling 
         constant $g_\rho^2$ and the neutron minus proton 
	 root mean square radius for $^{208}$Pb (in fm) 
	 are given along with the radii of 1/3, 1/2, 3/4 
	 and 1.4 $M_\odot$ neutron stars in km. Finally, 
	 the transition density $\rho_c$ between the inner 
	 crust and liquid interior is given in fm$^{-3}$.}
\begin{tabular} {lccccccc}
$\Lambda_{\rm v}$ & $g_\rho^2$ & $R_n-R_p$($^{208}$Pb) & 
R(${1\over 3}M_\odot$) & R(${1\over 2}M_\odot$) & 
R(${3\over 4}M_\odot$) & R($1.4M_\odot$) & $\rho_c$ \\
0.05 & 127.8389 & 0.1736 & 15.88 & 14.06 & 13.43&  13.25 & 0.0923 \\
0.04 & 116.2950 & 0.1895 & 17.75 & 15.00 & 13.96& 13.47 & 0.0866 \\
0.03 & 106.6635 & 0.2054 & 19.25 & 15.77 & 14.42& 13.65 & 0.0794 \\
0.02 & 98.5051 & 0.2215 & 19.80 & 16.24 & 14.76& 13.82 & 0.0717 \\
0.01 & 91.5061 & 0.2378 & 19.54 & 16.46 & 15.08& 14.07 & 0.0648 \\
0.00 & 85.4357 & 0.2543 & 18.95 & 16.53 & 15.43& 14.56 & 0.0594 \\
\label{Tablethree}
\end{tabular}
\end{table}

\begin{table}
\caption{Results for the Z271 parameter set with 
         $\Lambda_{\rm s}\!=\!0$. The $NN\rho$ coupling 
         constant $g_\rho^2$ and the neutron minus proton 
	 root mean square radius for $^{208}$Pb (in fm) 
	 are given along with the radii of 1/3, 1/2, 3/4 
	 and 1.4 $M_\odot$ neutron stars in km. Finally, 
	 the transition density $\rho_c$ between the inner 
	 crust and liquid interior is given in fm$^{-3}$.}
\begin{tabular} {lccccccc}
$\Lambda_{\rm v}$ & $g_\rho^2$ & $R_n-R_p$($^{208}$Pb) & 
R(${1\over 3}M_\odot$) & R(${1\over 2}M_\odot$) & 
R(${3\over 4}M_\odot$) & R($1.4M_\odot$) & $\rho_c$ \\
0.14 & 139.3368 & 0.1525 & 13.799 & 12.709 & 12.293 & 11.616 & 0.0974 \\
0.12 & 129.2795 & 0.1650 & 15.012 & 13.379 & 12.688 & 11.748 & 0.0936 \\
0.10 & 119.5245 & 0.1771 & 16.219 & 14.039 & 13.080 & 11.880 & 0.0890 \\
0.08 & 112.9710 & 0.1900 & 17.405 & 14.696 & 13.481 & 12.016 & 0.0845 \\
0.06 & 106.2682 & 0.2026 & 18.31 & 15.28 & 13.89 & 12.18 & 0.0796 \\
0.05 & 103.2065 & 0.2090 & 18.61 & 15.53 & 14.09 & 12.29 & 0.0771 \\
0.04 & 100.3162 & 0.2154 & 18.82 & 15.76 & 14.30 & 12.43 & 0.0747 \\
0.03 &  97.5834 & 0.2218 & 18.95 & 15.96 & 14.53 & 12.62 & 0.0725 \\
0.02 &  94.9956 & 0.2282 & 18.98 & 16.12 & 14.75 & 12.89 & 0.0703 \\
0.01 &  92.5415 & 0.2347 & 18.94 & 16.25 & 14.98 & 13.27 & 0.0683 \\
0.00 &  90.2110 & 0.2413 & 18.88 & 16.36 & 15.20 & 13.77 & 0.0665 \\
\label{Tablefour}
\end{tabular}
\end{table}
]

\twocolumn[\hsize\textwidth\columnwidth\hsize\csname
@twocolumnfalse\endcsname 

\begin{table}
\caption{Results for the Z271 parameter set with 
         $\Lambda_{\rm v}\!=\!0$. The $NN\rho$ coupling 
         constant $g_\rho^2$ and the neutron minus proton 
	 root mean square radius for $^{208}$Pb (in fm) 
	 are given along with the radii of 1/3, 1/2, 3/4 
	 and 1.4 $M_\odot$ neutron stars in km. Finally, 
	 the transition density $\rho_c$ between the inner 
	 crust and liquid interior is given in fm$^{-3}$.}
\begin{tabular} {lccccccc}
$\Lambda_{\rm s}$ & $g_\rho^2$ & $R_n-R_p$($^{208}$Pb) & 
R(${1\over 3}M_\odot$) & R(${1\over 2}M_\odot$) & 
R(${3\over 4}M_\odot$) & R($1.4M_\odot$) & $\rho_c$ \\
0.06 & 146.6988 & 0.1640 & 13.34 & 12.71 & 12.52 & 11.98 & 0.0844 \\
0.05 & 132.8358 & 0.1775 & 14.63 & 13.48 & 13.01 & 12.19 & 0.0830 \\
0.04 & 121.3666 & 0.1907 & 15.99 & 14.27 & 13.51 & 12.42 & 0.0807 \\
0.03 & 111.7205 & 0.2036 & 17.28 & 15.02 & 14.00 & 12.68 & 0.0777 \\
0.02 & 103.4949 & 0.2163 & 18.27 & 15.66 & 14.46 & 12.97 & 0.0741 \\
0.01 & 96.3974 & 0.2288 & 18.79 & 16.10 & 14.86 & 13.32 & 0.0702 \\
0.00 & 90.2110 & 0.2413 & 18.88 & 16.36 & 15.20 & 13.77 & 0.0665 \\ 
\label{Tablefive}
\end{tabular}
\end{table}
]

\section{Discussion and Conclusions}
\label{sec:conclusions}

A number of relativistic effective field theory parameter sets have
been used to study correlations between the radii of neutron stars and
the neutron radius $R_n$ of $^{208}$Pb. An RPA stability analysis was
employed to find the transition density between the nonuniform inner
crust and the uniform liquid interior. For the nonuniform outer crust
we invoked the EOS of Baym, Pethick, and Sutherland~\cite{baym}.
Then, a simple polytropic formula for the EOS, approximately valid for
most of the crust~\cite{link}, was used to interpolate between the
outer crust and the liquid interior. This simple, yet fairly accurate,
procedure allows us to study the EOS for a variety of parameter sets
that predict a wide range of values for the neutron radius of
$^{208}$Pb.

For a ``normal'' $1.4 M_\odot$ neutron star we find central densities
of several times normal nuclear matter saturation density
($\rho_0\!\approx\!0.15~{\rm fm}^3$). Because the neutron radius of
$^{208}$Pb does not constrain the high-density component of the EOS,
we find a strong model dependence between the radius of a $1.4M_\odot$ 
neutron star and $R_n$. In contrast, the central density of a low mass 
neutron star is close to $\rho_0$. Therefore properties of the low
mass star are sensitive to the EOS over the same density range as
$R_n$. As a result, we find a strong correlation and a weak model
dependence between the radius of a $0.5 M_\odot$ neutron star and 
$R_n$ (see Fig. \ref{Fig3}). Thus, it should be possible to infer 
some properties of low mass neutron stars from detailed measurements 
in atomic nuclei.

Understanding the density dependence of the equation of state is
particularly interesting. A softening of the EOS at high density 
(where the pressure rises slower than expected) could signal the 
transition to an exotic phase, such as pion/kaon condensates,
strange quark matter, and/or a color superconductor. Yet obtaining 
definitive results on the density dependence of the EOS may require 
measuring the radius of neutron stars for a broad range of masses.  
This may be difficult as most compilations to date find neutron
star masses in the very narrow range of 
$1.25-1.44~M_{\odot}$~\cite{masses}.
Further, important ambiguities will remain if the radius of a 
$1.4 M_\odot$ neutron star proves to be moderately small. Would
a small radius be an indication that the EOS is relatively soft 
at all densities and there is no phase transition, or is the 
EOS stiff at low density and undergoes an abrupt softening at
high density from a phase transition?

Therefore, it is important to also make measurements that are
exclusively sensitive to the low density EOS. One obvious possibility 
is the radius of a low mass ($\approx 0.5 M_\odot$) neutron star as 
it central density, close to $\rho_0$, is much smaller than that of a 
$1.4 M_\odot$ star. However such low mass stars may be very rare
because they are hard to form. Notably, the neutron radius of a heavy 
nucleus, such as $^{208}$Pb, contains similar information. Indeed, we
find a strong correlation and a weak model dependence between the 
neutron radius of $^{208}$Pb and the radius of a $0.5M_\odot$ neutron 
star. This allows one to use nuclear information to infer the radius 
of a low mass neutron star. Hence, comparing this inferred radius to 
the measured radius of a $\simeq\!1.4 M_\odot$ neutron star, should 
provide the most complete information to date on the density
dependence of the equation of state.

\smallskip
This work was supported in part by DOE grants DE-FG02-87ER40365 and
DE-FG05-92ER40750.

\twocolumn[\hsize\textwidth\columnwidth\hsize\csname
@twocolumnfalse\endcsname 
\section{Appendix}
The polarizations are defined in Eqs. (\ref{pis},\ref{pim},\ref{pil})
and describe particle-hole or particle antiparticle excitations. In
the static limit (energy transfer $q_0\!=\!0$) the scalar polarization
for momentum transfer $q$ is given by
\begin{equation}
\Pi_s(q,0)={1\over 2\pi^2}\left\{
          k_F E_F - \left(3M^{*2}+{q^2\over 2}\right) 
 	  \ln {k_F+E_F\over M^*}+ {2E_F E^2\over q}
	  \ln\Big|{2k_F-q\over 2k_F+q}\Big|-{2E^3\over q}
	  \ln\Big|{qE_F-2k_FE\over qE_F+2k_FE}\Big|\right\},
\end{equation}
with Fermi momentum $k_F$, nucleon effective mass $M^*$,  
$E_F=(k_F^2+{M^*}^2)^{1/2}$, and $E=(q^2/4+{M^*}^2)^{1/2}$.
Likewise, the longitudinal polarization is given by
\begin{equation}
\Pi_{00}(q,0)=-{1\over \pi^2}
       \left\{{2\over 3}k_F E_F - {q^2\over 6} 
       \ln {k_F+E_F\over M^*}- {E_F \over 3q}({M^*}^2+
       k_F^2-{3\over 4}q^2)\ln \Big|{2k_F-q\over 2k_F+q}\Big|
       +{E\over 3q}({M^*}^2-{q^2\over 2})
       \ln \Big|{qE_F-2k_FE\over qE_F+2k_FE}\Big|\right\},
\end{equation}
while the mixed scalar-vector polarization becomes
\begin{equation}
\Pi_m(q,0)={M^*\over 2\pi^2}
           \left\{k_F-({k_F^2\over q}-{q\over 4})
	   \ln \Big|{2k_F-q\over 2k_F+q}\Big|\right\}.
\end{equation}
]

\end{document}